\documentclass[aip,preprint,amsfonts,amsmath,amssymb]{revtex4-1}

\usepackage{graphicx}
\usepackage{bm}

\usepackage{color}
\usepackage{soul}

\DeclareUnicodeCharacter{2212}{-}

\usepackage{listings}

\begin{document}

\title{Effect of the metallicity on the capacitance of gold -- aqueous sodium chloride interfaces}
\author{Alessandra Serva}
\affiliation{Sorbonne Universit\'{e}, CNRS, Physico-chimie des \'Electrolytes et Nanosyst\`emes Interfaciaux, PHENIX, F-75005 Paris}
\affiliation{R\'eseau sur le Stockage Electrochimique de l'Energie (RS2E), FR CNRS 3459, 80039 Amiens Cedex, France}
\author{Laura Scalfi}
\affiliation{Sorbonne Universit\'{e}, CNRS, Physico-chimie des \'Electrolytes et Nanosyst\`emes Interfaciaux, PHENIX, F-75005 Paris}
\affiliation{R\'eseau sur le Stockage Electrochimique de l'Energie (RS2E), FR CNRS 3459, 80039 Amiens Cedex, France}
\author{Benjamin Rotenberg}
\affiliation{Sorbonne Universit\'{e}, CNRS, Physico-chimie des \'Electrolytes et Nanosyst\`emes Interfaciaux, PHENIX, F-75005 Paris}
\affiliation{R\'eseau sur le Stockage Electrochimique de l'Energie (RS2E), FR CNRS 3459, 80039 Amiens Cedex, France}
\author{Mathieu Salanne}
\email{mathieu.salanne@sorbonne-universite.fr}
\affiliation{Sorbonne Universit\'{e}, CNRS, Physico-chimie des \'Electrolytes et Nanosyst\`emes Interfaciaux, PHENIX, F-75005 Paris}
\affiliation{R\'eseau sur le Stockage Electrochimique de l'Energie (RS2E), FR CNRS 3459, 80039 Amiens Cedex, France}
\affiliation{Institut Universitaire de France (IUF), 75231 Paris Cedex 05, France}
%Add RS2E?

%%%%%%%%%%%%%%%%%%%%%%%%%%%%%%%%%%%%%%%%%%%%%%%%%%%%%%%%%%%%%%%%%%%%%%%%%%%%%%%%%%%%%%%%%%%%%

\begin{abstract}
Electrochemistry experiments have established that the capacitance of electrode-electrolyte interfaces is much larger for good metals such as gold and platinum than for carbon-based materials.
Despite the development of elaborate electrode interaction potentials, to date molecular dynamics simulations were not able to capture this effect. Here we show that changing the width of the Gaussian charge distribution used to represent the atomic charges in gold is an effective way to tune its metallicity. Larger Gaussian widths lead to a capacitance of aqueous solutions (pure water and 1~molar NaCl) in good agreement with recent {\it ab initio} molecular dynamics results. For pure water, the increase in the capacitance is not accompanied with structural changes, while in the presence of salt the Na$^+$ cations tend to adsorb significantly on the surface. For a strongly metallic gold electrode, these ions can even form inner sphere complexes on hollow sites of the surface.
\end{abstract}

\maketitle

%%%%%%%%%%%%%%%%%%%%%%%%%%%%%%%%%%%%%%%%%%%%%%%%%%%%%%%%%%%%%%%%%%%%%%%%%%%%%%%%%%%%%%%%%%%%%

\section{Introduction}

Classical molecular dynamics simulations of electrode -- electrolyte interfaces bring important microscopic insights for the interpretation of electrochemistry experiments~\cite{scalfi2021a}. They are not only relevant in the case of concentrated electrolytes, for which classical theories fail to capture many effects such as ion-ion correlations~\cite{kornyshev2007a}, but also in cases where the molecular structure of the solvent can impact the interfacial properties~\cite{son2021a}. 
Even though most of the recent works were dedicated to the study of supercapacitors~\cite{salanne2016a}, in which the electrode is generally a carbon material and the electrolyte an organic or ionic liquid, systems made of good metals such as platinum or gold put in contact with water were also considered. Such studies have evidenced very interesting phenomena, such as the formation of a strongly bound water adlayer at the surface of the metal displaying hydrophobic features~\cite{willard2009a,limmer2013b,willard2013a}. Recently, the role of this adlayer on the adsorption of hydrophobic solutes and on the energetics of electrochemical reactions occurring at the interface was further investigated~\cite{serva2021a}. However, these simulations were not able to reproduce one of the key experimental quantities, namely the capacitance of the interface. Extracting quantitative numbers from experiments is difficult due to many specific adsorption processes, but it is well accepted that this value should typically be in the range of 20 to 50~$\mu$F~cm$^{-2}$, although recent experiments on platinum suggest even larger values~\cite{ohja2020a}, up to 100--200~$\mu$F~cm$^{-2}$. So far, simulation studies only reported single-electrode capacitances in the range of 5--10~$\mu$F~cm$^{-2}$ for aqueous electrolytes~\cite{willard2009a}. 

Although such lower capacitance could be attributed to specific interactions occurring at the interface, including chemisorption effects that would not be captured by conventional force fields, recent {\it ab initio} molecular dynamics simulations of interfaces with gold electrodes do not support this hypothesis~\cite{goldsmith2021a,li2019b}. Indeed these calculations yield estimates of the capacitance of 15~$\mu$F~cm$^{-2}$ to 30~$\mu$F~cm$^{-2}$, while the structure of the adsorbed water layer is very similar to the one of previous classical simulations. This therefore points towards a limitation of the classical models used in the molecular dynamics to describe the metal and its interactions with the electrolyte. In particular, its metallicity, or metallic character, reflects the strength of the electronic response to external fields such as the ones due to the adsorption of charged or polarized species, and is expected to play an important role. In a recent work, we have introduced a semi-classical Thomas-Fermi electrode model to account for the kinetic energy of the electrons. This allowed us to study the effect of the Thomas-Fermi screening length on the accumulation of charge at the electrode surface, the structure of the interfacial electrolyte and the capacitance in conditions corresponding to materials with poor metallic properties such as carbon~\cite{scalfi2020b}. In parallel, Sato and co-workers have shown that similar effects can be expected by changing the width of the Gaussian charge distributions which are used for the electrode atoms~\cite{nakano2019a,oshiki2021a}. Here we build on these works to investigate the impact of the latter parameter in the case of gold electrodes put in contact with aqueous solutions. We show that increasing the Gaussian width may lead to a substantial increase of the capacitance, which is accompanied by changes of the ionic adsorption profiles, while the water layers remain largely unchanged.  

\section{Choice of the electrode model parameters}

Over the past decade, fluctuating-charge models have been widely used to simulate electrochemical systems in which the potential of electrode atoms is fixed. First introduced by Siepmann and Sprik~\cite{siepmann1995a}, the main ingredients are (i) the use of atom-centered Gaussian charge distributions for the electrode atoms: 
\begin{equation}
\rho_i({\bf r})=q_i (2\pi\sigma^2)^{-3/2}e^{-\mid{\bf r}-{\bf r}_i\mid^2 / 2 \sigma^2}	
\end{equation}
\noindent where $\sigma$ is the width of the distribution and $r_i$ is the position of atom $i$, and (ii) the application of a constant potential condition for all the atoms within an electrode material. As discussed at length in Ref.~\citenum{nakano2019a}, the use of the Gaussian charges introduces a self-interaction term inversely proportional to $\sigma$, which can thus be linked to the hardness parameter $H$ of the charge-equilibration method~\cite{mortier1986a,rappe1991a} as $H~\sim~1/\sigma\sqrt{\pi}$. The latter is generally used in other contexts, but it was recently used in a series of works devoted to electrode-electrolyte interfaces~\cite{onofrio2015a,onofrio2015b,buraschi2020a}. Note that besides this self-energy term, the Gaussian width also impacts all the electrostatic interactions involving electrode atoms so that the two approaches are not strictly equivalent.

It is thus instructive to compare the value of the hardness corresponding to the usual choice for the Gaussian width with typical hardness values from the literature. Siepmann and Sprik proposed to adjust this value in order to match the energy of a system made of a single charge adsorbed on a flat metallic surface to the one obtained in the continuum limit, yielding $\sigma = 0.40$~\AA\ for platinum~\cite{siepmann1995a}. In the absence of a generic criterion, this value was then also used for other materials. It is worth noting that most of these studies focused on carbon and generally yielded interfacial properties in good agreement with experiments. Yet, as discussed in Ref.~\citenum{oshiki2021a}, this corresponds to a hardness $H\approx$~20~eV, quite far from the expected gas-phase value for carbon (10~eV), gold (6.92~eV) or platinum (7~eV)~\cite{pearson1988a}. Even though atomic parameters such as hardness are not expected to be directly transferable from gaseous to condensed phase, because of the different electronic structures, such a gap calls for an examination of the influence of this quantity.

We recently introduced a semi-classical Thomas-Fermi electrode model, with an additional term accounting for the kinetic energy of the electrons in the metal~\cite{scalfi2020b}. This term, arising from a description of the electronic properties of the materials instead of the atoms, contributes in the classical description to the self-energy of the electrode atoms by increasing the ``effective'' hardness. We analyzed the impact of this screening length at fixed Gaussian width $\sigma = 0.40$~\AA, which corresponds to materials with poor metallic character. In this work, we focus instead on a good metal, gold, and study the effect of increasing the Gaussian width, from an initial value of 0.43~\AA\ to an upper limit of 1.17~\AA, which corresponds to the gas-phase hardness of the atom. 

\section{Simulation details}

\begin{figure*}[hbt!]
\centering
\includegraphics[width=0.8\textwidth]{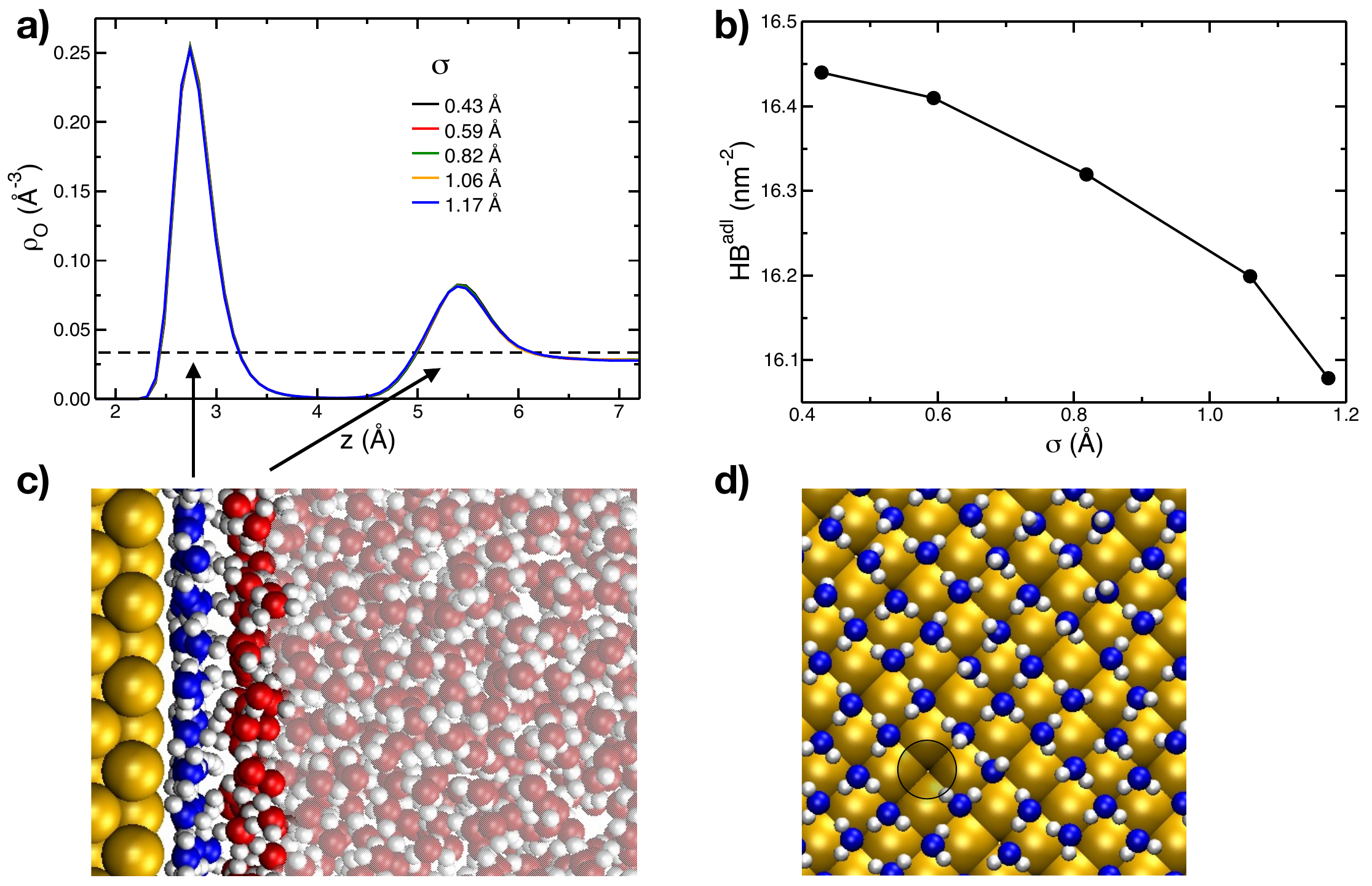}
	\caption{a) Atomic density profiles for the O atoms near the electrode at null potential for Gaussian widths $\sigma$ varying from 0.43 to 1.17~\AA. b) Hydrogen bond density within the adlayer water molecules as a function of the Gaussian width. c) Representative snapshot of the system (lateral view), with Au atoms in yellow,  O atoms of the water molecules of the adlayer in blue, O atoms of the other water molecules in red and H atoms in white. d) Representative snapshot of the system (top view of the adlayer, which shows a square symmetry with some vacancies); the same coloring scheme is used for the atoms.}
\label{fig:h2o}
\end{figure*}

All simulations were performed using the molecular dynamics code MetalWalls~\cite{marinlafleche2020a}. Two different systems were simulated, both with two gold electrodes, each with 1620 atoms distributed in 10 planes and their (100) face in contact with the liquid between them. In the first case, the liquid consists of pure water (2160 molecules), while in the second it also contains 39 NaCl pairs, resulting in a 1~mol~L$^{-1}$ NaCl$_{(aq)}$ electrolyte. In both cases, the lateral box dimensions are $L_x$~=$L_y$~=36.63~\AA.

For each system, the distance between the two inner planes of the electrodes was equilibrated at atmospheric pressure by applying a constant pressure force to the electrodes, yielding distances of 48.93~\AA\, %50.67~\AA\ 
and 50.67~\AA\ for the two systems, respectively. As in our previous work~\cite{scalfi2020b}, water molecules are modeled using the SPC/E force field~\cite{berendsen1987a}, while the Lennard-Jones parameters for Na$^+$ and Cl$^-$ were taken from Ref.~\citenum{dang1995c} and the ones for the electrode atoms from Ref.~\citenum{berg2017a}. The simulations were run in the $NVT$ ensemble with a timestep of 2~fs; a Nos\'e-Hoover thermostat chain~\cite{martyna1992a} was used with a time constant of 1~ps to maintain a temperature of 298~K. The electrode charges were calculated at each timestep using a matrix inversion method~\cite{scalfi2020a} to enforce both a constant applied potential of 0~V betweeen the two electrodes and
the electroneutrality constraints on the charges. Widths of 0.43, 0.59, 0.82, 1.06 and 1.17~\AA\ were used for the
Gaussian charge distribution. In the case of pure water the simulation times used for equilibration and production were of 1~ns and 5~ns. For the second system, they were of 1~ns and 9~ns except for the largest value of $\sigma$ (1.17~\AA) for which they were increased to 2~ns and 25~ns in order to reach convergence for the structural properties and for the calculation of the differential capacitance.

\section{Gold -- water interface}

We first focus on the most simple system in which the gold electrodes are separated by pure water. Figure \ref{fig:h2o}a shows that the atomic density profiles of water across the electrochemical cell do not change with the Gaussian width. The structure at the interface was already described in previous work~\cite{clabaut2020a}: Water molecules form a tightly bound adlayer (located at less  than 4.0~\AA\ from the surface), with an average of 12 molecules per nm$^2$, in which they adopt a flat orientation -- see the snapshots of a lateral and a top view of this adlayer on Figures \ref{fig:h2o}c and \ref{fig:h2o}d. To further characterize the structural changes, the number of hydrogen bonds (HB) per unit area  was determined within these adlayer water molecules. Two molecules are considered to form a HB~\cite{white2000a} when the O(-H)$\cdots$O distance is smaller than 3.2~\AA\ and the corresponding $\widehat{\rm OHO}$ angle is in the range $[140^\circ-220^\circ]$.

\begin{figure}[hbt!]
\centering
\includegraphics[width=0.5\columnwidth]{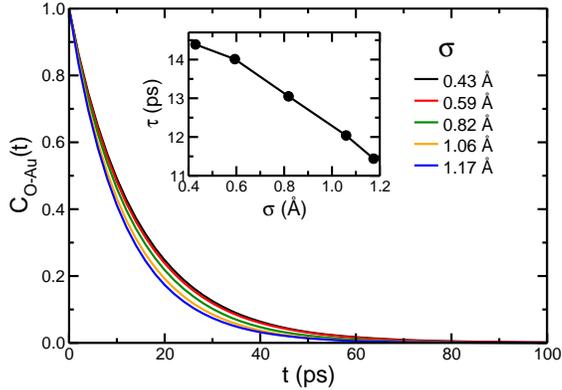}
	\caption{Time-correlation functions for the adsorption of water molecules within the adlayer for Gaussian widths varying from 0.43 to 1.17~\AA\ (the corresponding characteristic adsorption lifetimes are provided in the inset).}
\label{fig:lifetimeh2o}
\end{figure}

Figure ~\ref{fig:h2o}b shows that the number of HBs within the adlayer decreases with the Gaussian width: On average there are 16.55~HBs~nm$^{-2}$ for $\sigma$~=~0.43~\AA\ and 16.05~HBs~nm$^{-2}$ for $\sigma$~=~1.17~\AA. This suggests that the water molecules may be less bonded within the adlayer and less strongly bound to the surface. To  investigate this possibility, we estimate the characteristic adsorption lifetime of the water molecules, $\tau$, as the time integral of the following correlation function:
\begin{equation}
	C_{{\rm O-Au}}(t)=\frac{\langle h(0)h(t)\rangle}{\langle h(0)h(0)\rangle}
\end{equation}
\noindent where $h(t)=1$ if a water molecule is continuously interacting with the gold surface from time 0 to time $t$, and 0 otherwise. A water molecule is considered as interacting with the Au surface via its O atom if: (1) the distance between the inner Au plane and the water oxygen atom is less than 4.0 \AA~(\emph{i.e.} the water molecule belongs to the adlayer) and (2) both Au-O-H angles are greater than 60$^\circ$. This characteristic adsorption lifetime decreases from 14 to 11~ps when the gold charges Gaussian width increases (Figure ~\ref{fig:lifetimeh2o}).

These structural and dynamic changes remain relatively small, in agreement with our previous work on poorly metallic electrodes~\cite{scalfi2020b}. Nevertheless, the latter study showed that the capacitance is generally more sensitive to the electrode metallicity, so that we determined the differential capacitance from the fluctuations of the total charge of the electrode~\cite{scalfi2020a}:
\begin{equation}
	\mathcal{C}_{\rm diff}=\beta \langle \delta Q_L^2 \rangle + \mathcal{C}_{\rm diff}^{\rm empty} \,,
\end{equation}
\noindent where $\beta=1/k_B T$, $\delta Q_L = Q_L-\langle Q_L \rangle$ with $Q_L$ the instantaneous total charge of the left electrode (which is strictly opposite to the charge of the right electrode since an electroneutrality constraint is imposed at each timestep) and $\mathcal{C}_{\rm diff}^{\rm empty}$ is a contribution arising from the charge fluctuations suppressed in the Born-Oppenheimer dynamics of the charges and corresponding to the capacitance of the empty capacitor~\cite{scalfi2020a} (\emph{i.e.} in the absence of liquid between the electrodes). Note that $\mathcal{C}_{\rm diff}$ is  the capacitance of the full system. Since there is no voltage between the two electrodes, they are equivalent and the single-electrode capacitance is simply $C_{\rm diff}=2\mathcal{C}_{\rm diff}$. 

\begin{figure}[hbt!]
\centering
\includegraphics[width=0.5\columnwidth]{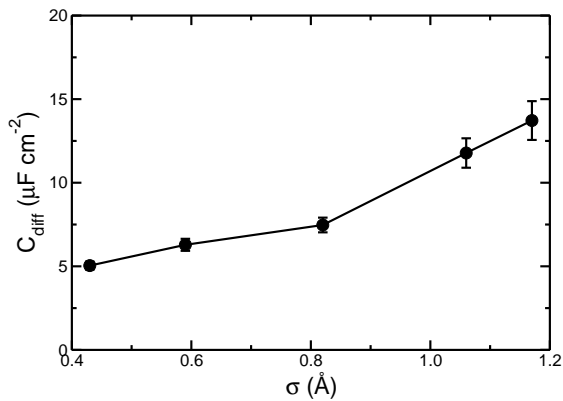}
	\caption{Single-electrode differential capacitance as a function of the Gaussian width for the pure water system. Error bars correspond to 95~\% confidence intervals.}
\label{fig:capacitanceh2o}
\end{figure}

The capacitance is shown as a function of the Gaussian width on Figure \ref{fig:capacitanceh2o}, with confidence intervals  obtained from the standard error on the variance corrected for data correlations as $\langle \delta Q_L^2\rangle \times \sqrt{4\tau_C/\tau_S}$ where $\tau_S$ is the total sampling time and $\tau_C = \int_0^\infty \langle \delta Q_L(t)\delta Q_L(0)\rangle^2/\langle \delta Q_L^2\rangle^2 {\rm d}t$ is the correlation time~\cite{zwanzig1969a}. There is a clear effect of the Gaussian width since the capacitance switches from 5~$\mu$F~cm$^{-2}$ for $\sigma$~=~0.43~\AA, in agreement with previous works,~\cite{willard2009a} to a value of 13~$\mu$F~cm$^{-2}$  for $\sigma$~=~1.17~\AA. Due to the insulating character of pure water, it is not possible to compare these numbers with an experimental value. However recent {\it ab initio} molecular dynamics simulations performed on (111) surfaces of gold provided capacitances of 15 and 30~$\mu$F~cm$^{-2}$. The difference between the two {\it ab initio} values probably lies in the different setups. Indeed, the first work involves a single electrode and the water layer consists of a slice of explicit water molecules followed by a dielectric continuum,~\cite{goldsmith2021a} while in the second one the electrochemical cell is made of a slab of gold in contact with a slab of pure water molecules only.~\cite{li2019b} The method for charging the electrodes was also different since in the first case a net charge was added to the system, while in the second case it was achieved by introducing explicit Na$^+$ counterions. In addition, due to the use of density functional theory to calculate the atomic forces both studies were limited in terms of system size and sampling time, with production trajectories of 10~ps only. Nevertheless, we note that the capacitance we obtain for the larger Gaussian width is in much better agreement with these two studies than the prediction of the more usual, smaller width.

\section{Gold -- aqueous sodium chloride interface}

\begin{figure*}[hbt!]
\centering
\includegraphics[width=0.8\textwidth]{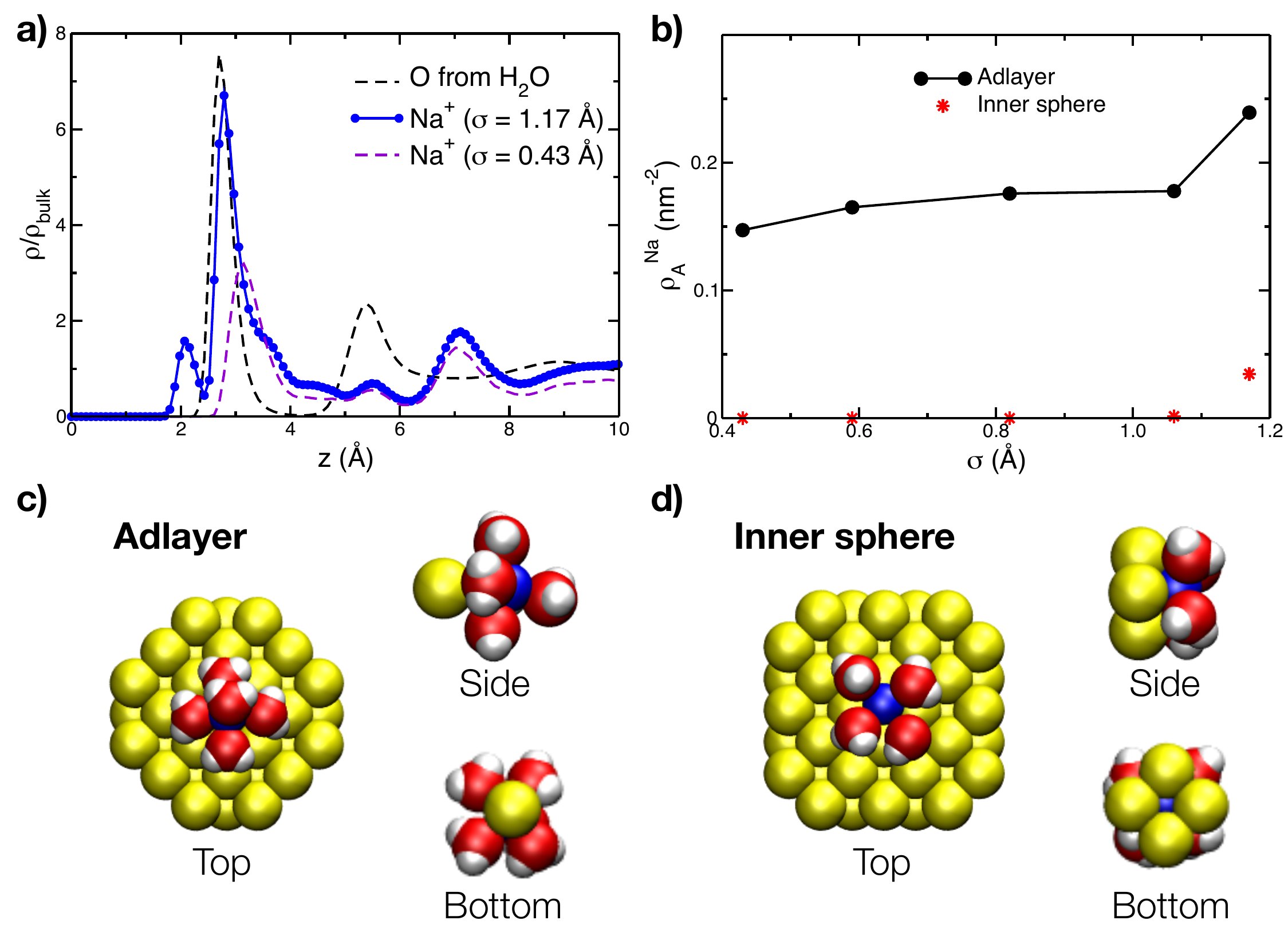}
	\caption{a) Atomic density profiles for the Na$^+$ ions near the electrode at null potential for Gaussian widths of 0.43 and 1.17~\AA. The density profile of oxygen atoms is also shown for comparison. b) Surfacic concentration of the Na$^+$ ions in the two identified adsorption sites. c) Typical snapshot of a Na$^+$ ion adsorbed within the water adlayer (2.45~\AA~$< z <$~4.05~\AA). d) Typical snapshot of a Na$^+$ ion adsorbed in an inner sphere site ($z < 2.45$~\AA).}
\label{fig:structurena}
\end{figure*}

The structural variation is much more pronounced for the aqueous NaCl solution. Indeed, although the adsorption profile of the water molecules also remains unchanged, it is not the case for the ions, in particular for the Na$^+$. As shown on Figure \ref{fig:structurena}a, and in agreement with our previous study,~\cite{scalfi2020b} the first layer of Na$^+$ ions is located at approximately 3.1~\AA\ from the surface for the smallest Gaussian width, \emph{i.e.} they lie just above the water adlayer. The corresponding concentration is of 0.15 ions nm$^{-2}$ as shown on Figure \ref{fig:structurena}b. Further examination of the structure of this adsorption mode leads to the picture shown on Figure \ref{fig:structurena}c: The ion is located on top of a gold atom; it is solvated by 3.5 water molecules from the adlayer on average (4 on the snapshot shown) and by 1 water molecule from the second water layer, which corresponds to a typical octahedral coordination shell (consisting of 1 gold atom and 5 water molecules in total). When the Gaussian width is increased, the preferential position of Na$^+$ ions progressively shifts inside the water adlayer; as shown on the Figure~\ref{fig:structurena}a the corresponding peak is almost superimposed with the one of the oxygen atoms. In parallel, their surfacic concentration increases, slightly until $\sigma$~=~1.06~\AA\ and more noticeably for $\sigma$~=~1.17~\AA. 

\begin{figure}[hbt!]
\centering
\includegraphics[width=0.5\columnwidth]{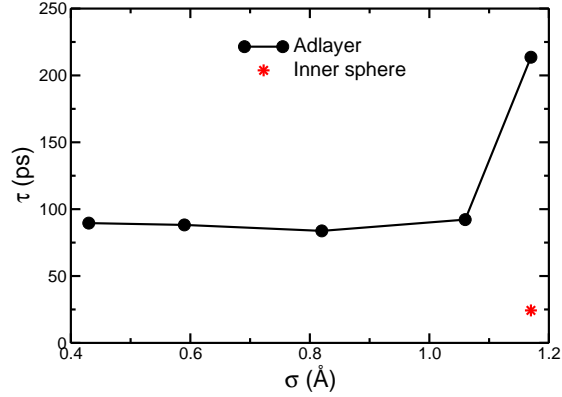}
	\caption{Characteristic adsorption lifetimes of Na$^+$ ions in the two adsorption sites as a function of the Gaussian width.}
\label{fig:lifetimenacl}
\end{figure}

For the latter width, we also observe the appearance of another adsorption mode in which the Na$^+$ ions adopt an inner sphere structure. It is represented on Figure \ref{fig:structurena}d, where it appears that the ion sits in a hollow site of the gold surface and is thus coordinated to four gold atoms. In addition, it remains solvated by 4 water molecules from the adlayer. The distance between the Na$^+$ ions and the surface in this adsorption mode, which is of approximately 2.1~\AA, is in good agreement with previous DFT calculations made on similar surfaces,~\cite{geada2018a} and the corresponding concentration is of 0.07 ions nm$^{-2}$. The calculated Na$^+$ adsorption lifetimes are provided on Figure \ref{fig:lifetimenacl}. They are larger by almost one order of magnitude than the ones previously calculated for water molecules for most of the Gaussian widths ($\tau \approx 90$~ps) except for the largest one, for which we observe a large increase for the ions adsorbed in the adlayer. The lifetime of the Na$^+$ ions in the inner sphere is noticeably shorter (25~ps).

\begin{figure*}[hbt!]
\centering
\includegraphics[width=0.8\textwidth]{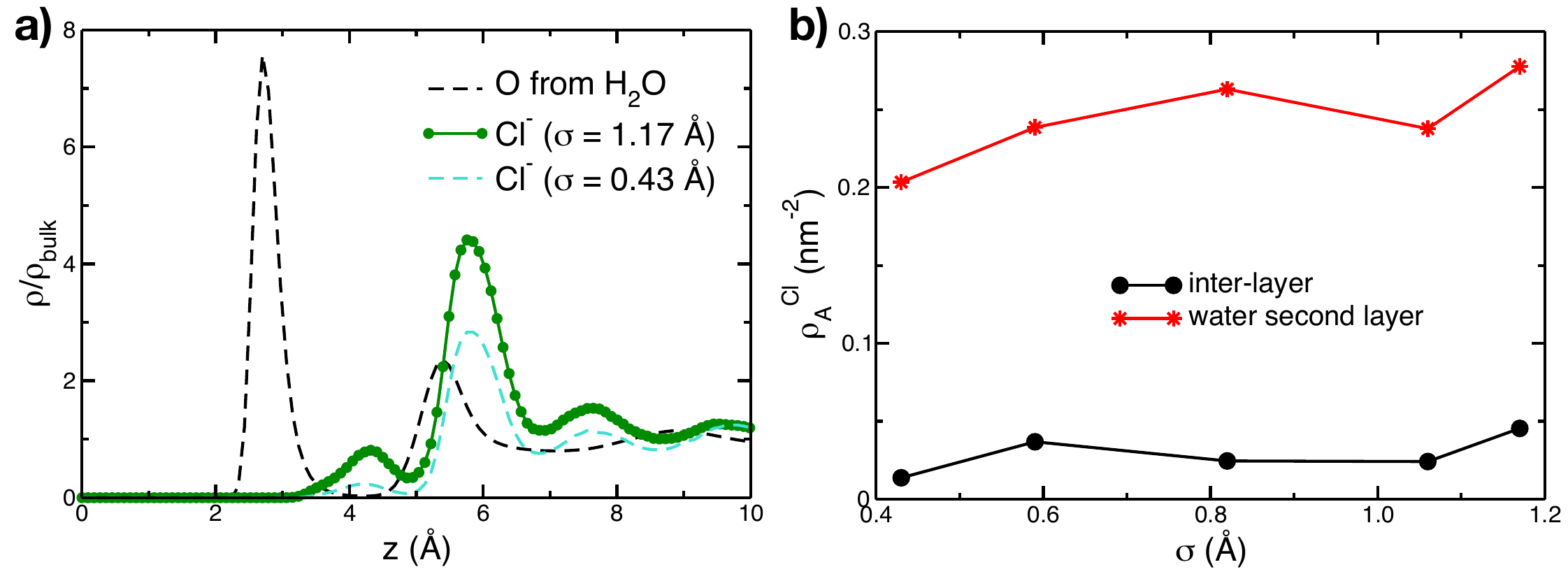}
	\caption{a) Atomic density profiles for the Cl$^-$ ions near the electrode at null potential for Gaussian widths of 0.43 and 1.17~\AA. The density profile of oxygen atoms is also shown for comparison. b) Surfacic concentration of the Cl$^-$ ions in the two identified adsorption sites.}
\label{fig:structurecl}
\end{figure*}

The adsorption profile of the chloride anions is very different from the one of the sodium cations, due to their larger ionic radius and to their different solvation shell structure. As shown on Figure \ref{fig:structurecl}a, the Cl$^-$ ions also have two different adsorption positions, but in that case they are both observed for all the $\sigma$ values. In the first adsorption mode they sit in the inter-layer region located between the water adlayer and the bulk liquid, and in the second one they are located slightly below the water second layer. The corresponding concentrations increase progressively with the Gaussian width (Figure \ref{fig:structurecl}b), but the variation is much weaker than in the case of the Na$^+$ ions.

\begin{figure}[hbt!]
\centering
\includegraphics[width=0.5\columnwidth]{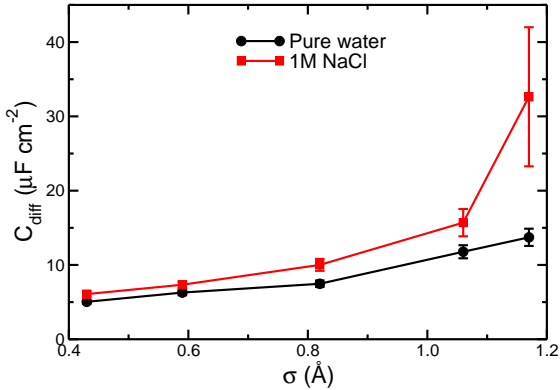}
	\caption{Single-electrode differential capacitance with respect to the Gaussian width for the aqueous NaCl electrolyte. Error bars correspond to the 95~\% confidence interval.}
\label{fig:capacitancenacl}
\end{figure}

The increase of the capacitance with the Gaussian width is much more pronounced in the presence of salt than in pure water, as shown on Figure \ref{fig:capacitancenacl}. The value obtained for the largest Gaussian width (33~$\mu$F~cm$^{-2}$) is much closer to experimental results generally obtained with good metals such as gold and platinum than in previous simulation studies. It is also worth noting it agrees well with the {\it ab initio} MD simulation in which Na$^+$ counterions were explicitly included.~\cite{li2019b} It is therefore possible that the discrepancy observed between the two previously discussed {\it ab initio} studies is due to the effect of ion adsorption. Increasing the Gaussian width to 1.25~\AA, which is the largest value for which simulations remained stable (with respect to the self-consistent calculation of electrode charges), further confirmed the trend since it yielded a capacitance as large as 125~$\mu$F~cm$^{-2}$, but with a very wide confidence interval. In addition, 50~ns production runs were not sufficient to converge the adsorption lifetimes of the ions. Although such a capacitance seems to be too large, we note that recent experimental work reported values of the same order of magnitude~\cite{ohja2020a}. This suggests that such large $\sigma$ values may deserve more attention in the future.

\begin{figure*}[hbt!]
\centering
\includegraphics[width=0.8\textwidth]{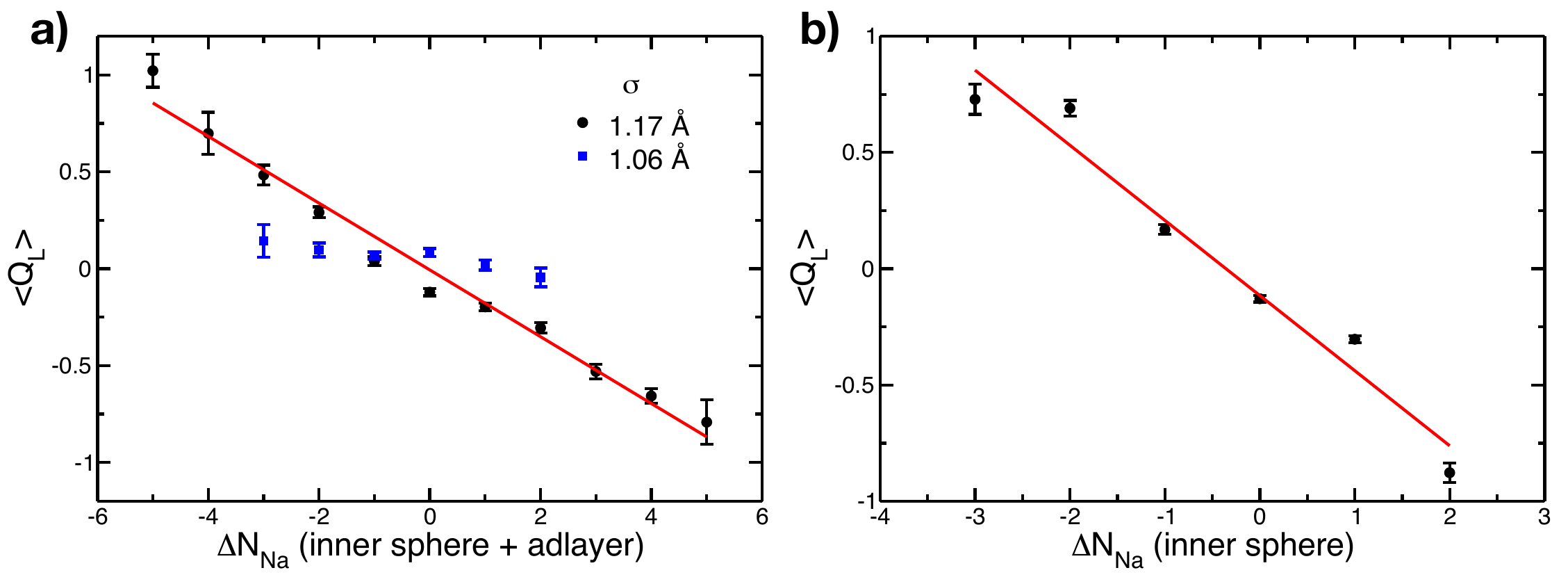}
	\caption{Average charge on the left electrode as a function of the imbalance in the number of adsorbed cations between the left and right electrodes (black and blue symbols: simulation data; red line: linear regressions for the $\sigma$~=~1.17~\AA\ case). a) All the Na$^+$ ions in both the adlayer and in inner sphere positions are considered b) Only the Na$^+$ ions in inner sphere positions are considered. Error bars correspond to 95~\% confidence intervals.}
\label{fig:correlationchargenaadsorption}
\end{figure*}
In order to assess the role of the ion adsorption/desorption on the electrode charge fluctuations, hence on the capacitance, we have computed the average charge on the left electrode at fixed Na$^+$ imbalance $\Delta N_{\rm Na}=N_{\rm Na, left}-N_{\rm Na, right}$ between the two electrodes, where $N$ can be computed either taking into account all the adsorbed layer or restricting to the inner sphere only. In the former case, results are shown for the two largest $\sigma$ values on Figure \ref{fig:correlationchargenaadsorption}a. For $\sigma$~=~1.06~\AA\ no correlation is visible, while for 1.17~\AA\ the two quantities are anti-correlated, reflecting the opposite charge induced by an adsorbed ion. In the absence of any other species, one would expect $\langle Q_L \rangle$ to scale as -0.5~$\Delta N$. Here we obtain a slope of -0.17, which points to additional mechanisms, such as changes in the interfacial chloride density or the reorganization of the water molecules. When the analysis is restricted to the inner sphere, the anti-correlation is also visible, with a larger slope of -0.32, showing the crucial role played by these ions on the overall capacitance.

\section{Conclusion}
We have evidenced the importance of the parameterization of the electrostatic parameters on the electrode/electrolyte properties. More precisely, changing the width of the Gaussian charge distribution used to represent the atomic charges in the electrode is an effective way to tune its metallicity. By focusing on aqueous interfaces with gold electrodes, we have shown that larger Gaussian widths lead to an increase in the interfacial capacitance -- for the considered nanocapacitors, by a factor 3 in the case of pure water, and even larger for 1~molar NaCl. The values obtained for the largest Gaussian widths are in much better agreement (compared to the typical width used in the literature) with experimental data available as well as with recent {\it ab initio} molecular dynamics results. For pure water, the  evolution of the capacitance with the Gaussian width is not accompanied with structural changes, while in the presence of salt the Na$^+$ cations tend to adsorb significantly on the surface. In particular, we observe the appearance of an inner sphere adsorption mode on hollow sites of the surface. These results open the way towards a more quantitative comparison between interfaces of metallic electrodes and varying electrolytes. For example, it will be very interesting to study the impact of the metallicity on highly concentrated electrolytes and ionic liquids.

\section*{Acknowledgements}
The authors thank J. Linnemann and K. Tschulik for useful discussions. This project has received funding from the European Research Council  under the European Union's Horizon 2020 research and innovation programme (grant agreements No. 771294 and 863473). This work was supported by the French National Research Agency (Labex STORE-EX, Grant  ANR-10-LABX-0076). The authors acknowledge HPC resources granted by GENCI (resources of IDRIS, Grant No. A0100910463) and by the HPCaVe Centre at Sorbonne Universit\'e. 

%\bibliography{references}

%merlin.mbs aipnum4-1.bst 2010-07-25 4.21a (PWD, AO, DPC) hacked
%Control: key (0)
%Control: author (8) initials jnrlst
%Control: editor formatted (1) identically to author
%Control: production of article title (0) allowed
%Control: page (1) range
%Control: year (1) truncated
%Control: production of eprint (0) enabled
%
\end{document}